\documentclass[11pt,a4paper]{article}
\usepackage{jcappub}
\title{Spin precession in anisotropic cosmologies}
\author[a,b]{A.Yu. Kamenshchik}
\author[c]{O.V. Teryaev}
\affiliation[a]{Dipartimento di Fisica e Astronomia, Universit\`a di Bologna and INFN,\\ via Irnerio 46, 40126 Bologna, Italy}
\affiliation[b]{L.D. Landau Institute for Theoretical Physics,\\ Kosygin str. 2,
119334 Moscow, Russia}
\affiliation[c]{Bogoliubov Laboratory of Theoretical Physics, Joint Institute for Nuclear Research,  141980 Dubna, Russia}
\affiliation[d]{Lomonosov Moscow State University,GSP-1, Leninskie Gory, 119991 Moscow, Russia} 
\emailAdd{Alexander.Kamenshchik@bo.infn.it}
\emailAdd{teryaev@theor.jinr.ru}
\abstract{We consider the precession of a Dirac particle spin in some anisotropic Bianchi universes.
This effect is present already in the Bianchi-I universe. We discuss in some detail  the geodesics and the spin precession for both  the Kasner and the Heckmann-Schucking solutions.  In the Bianchi-IX universe the spin precession  acquires the chaotic character due to the stochasticity of the oscillatory approach to the cosmological singularity. The related helicity flip of fermions in  the very early Universe may produce the 
sterile particles contributing to dark matter. 
}
\begin{document}
\maketitle
\flushbottom

\section{Introduction}

In almost all the  applications of  mathematical cosmology to the elaboration of observational data the isotropic Friedmann cosmological models are used. However, in the very early universe, the effects of anisotropies could be essential. As is well known, the most simple and well studied anisotropic cosmological models are the spatially homogeneous Bianchi models
(see e.g. \cite{LL,Ryan}). Remarkably, already in Bianchi models one can observe such interesting and important phenomenon as the oscillatory approach to the cosmological singularity \cite{BKL,Misner}. However, up to our knowledge, the behavior of quantum particles in the Bianchi universes was not studied in detail. We think that the filling of this gap can be of interest not only from the theoretical point of view, but can also  reveal some interesting physical effects in the very early universe. 

Especially promising can be study of motion of Dirac particles (quarks and leptons) in  gravitational fields. While this study has rather a long history \cite{old}, some essential progress was achieved in a recent series of papers \cite{ter-ob-sil,ter-ob-sil1,ter-ob-sil5}. In particular, the general expressions, characterising the spin motion in rather general gravitational fields were elaborated in paper \cite{ter-ob-sil5}. Here we apply this formalism to the study of the behavior of quantum particles with spin in some Bianchi universes. We found a novel effect  
of anisotropy induced spin precession, revealed already in the simplest case of the Bianchi-I universe.   We consider in some details the geodesics and the spin precession in Bianchi-I universes, making a special emphasis on the Kasner \cite{Kasner} and the Heckmann-Schucking 
\cite{Heck-Schuck}  solutions.   It is interesting also from the point of view of study of cosmic jets, which was undertaken in paper \cite{Mashhoon1}. 

Then we consider the precession in the Bianchi-IX universe.  Here, first of all, two qualitatively different contributions to the angular velocity are present and, second, the oscillatory approach to the singularity \cite{BKL} implies  the stochasticity of the changes of the direction of the precession axis.  

We also consider the possible physical consequences of these effects in very early Universe, including the appearance of effective magnetic field. 
The  similar precession effects are also present for classical rotators due to the  equivalence principle and  might be manifested in the structure formation in the very early universe. 

The equivalence principle implied also the helicity flip which is of special interest for massive Dirac neutrinos. 
The neutrinos produced as active ones are becoming sterile due to gravity-induced helicity flip and 
may contribute to fermionic dark matter.  
The structure of the paper is as follows: in the second section we briefly describe the precession of the Dirac particle in gravitational field; in the third section we give the general formulae for geodesics and spin rotation in the Bianchi-I universes, and, in particular, in the empty Bianchi-I universes evolving, following the Kasner solution; the section 4 is devoted to 
the Heckmann-Schucking solution for a Bianchi-I universe filled with a dust-like matter; in the fifth section we consider the precession of spin in a Bianchi-IX universe; in the concluding section we discuss possible physical applications of described effects and give a short outlook of the future directions of investigations. 

 \section{The precession of the Dirac particle in a gravitational field}

 Following the paper \cite{ter-ob-sil5}, we present the general formula for the precession of the Dirac particle, adapted for the case of Bianchi universes, where we shall use the sinchroneous reference frame.  
The metric can be presented as 
\begin{equation}
ds^2 = dt^2 - \delta_{\hat{a}\hat{b}}
W_{c}^{\hat{a}} 
W_{d}^{\hat{b}}dx^{c}dx^{d},
\label{metric}
\end{equation}
where $a,b,\cdots$ are world spatial indices, while the ones with the hats are spatial tetrad indices.  
We shall introduce also the inverse matrix $W_{\hat{c}}^a$ such that 
$W_{\hat{c}}^aW_{b}^{\hat{c}}=\delta_b^a$.

In paper \cite{ter-ob-sil5} it was shown that the average spin $\vec{s}$ in the semiclassical approximation is precessing with an angular velocity $\vec{\Omega}$ like 
\begin{equation}
\frac{d\vec{s}}{dt}=\vec{\Omega}\times\vec{s}=(\vec{\Omega}_{(1)}+\vec{\Omega}_{(2)})\times \vec{s}.
\label{prec}
\end{equation}
The velocities $\vec{\Omega}_{(1)}$ and $\vec{\Omega}_{(2)}$ correspond to gravitoelectric and to gravitomagnetic forces respectively.  
Then,
\begin{equation}
\Omega^{\hat{a}}_{(1)}=\frac{1}{\varepsilon'}W_{\hat{c}}^{d}p_{d}
\left
(\frac12
\Upsilon \delta^{\hat{a}\hat{c}}-\varepsilon^{\hat{a}\hat{e}\hat{f}}
C_{\hat{e}\hat{f}}^{\hat{c}}\right),
\label{prec1}
\end{equation}
\begin{equation}
\Omega^{\hat{a}}_{(2)}=\frac12\Xi^{\hat{a}}-\frac{1}{\varepsilon'(\varepsilon'+m)}\varepsilon^{\hat{a}\hat{b}\hat{c}}Q_{(\hat{b}\hat{d})}\delta^{\hat{d}\hat{n}}W_{\hat{n}}^kp_kW_{\hat{c}}^{l}p_l.
\label{prec2}
\end{equation}
Here $C_{\hat{a}\hat{b}}^{\hat{c}}$ are anholonomity coefficients 
\begin{equation}
C_{\hat{a}\hat{b}}^{\hat{c}}=W_{\hat{a}}^dW_{\hat{b}}^e
\partial_{[d}\dot{W}_{e]}^{\hat{c}},\ C_{\hat{a}\hat{b}\hat{c}}=g_{\hat{c}\hat{d}}C_{\hat{a}\hat{b}}^{\hat{d}}.
\label{prec3}
\end{equation}
Then, 
\begin{equation}
Q_{\hat{a}\hat{b}}=g_{\hat{a}\hat{c}}W_{\hat{b}}^d\dot{W}_{d}^{\hat{c}},
\label{Q}
\end{equation}
\begin{equation}
\Upsilon = -\varepsilon^{\hat{a}\hat{b}\hat{c}}C_{\hat{a}\hat{b}\hat{c}},
\label{Up}
\end{equation}
\begin{equation}
\Xi_{\hat{a}}=\varepsilon_{\hat{a}\hat{b}\hat{c}}Q_{\hat{b}\hat{c}}.
\label{Xi}
\end{equation}

The motion of the particle is characterised by its momentum $p_a$ and by the energy
$\varepsilon'=\sqrt{m^2+\delta^{\hat{c}\hat{d}}W_{\hat{c}}^aW_{\hat{d}}^bp_ap_b}$.
It can be absorbed together with the particle mass $m$, its momentum $p_a$   by introducing the velocity $v_a$. Thus, the precession velocities are 
\begin{equation}
\Omega^{\hat{a}}_{(1)}=v_{\hat{c}}\left(\frac12\Upsilon\delta^{\hat{a}\hat{c}}-\varepsilon^{\hat{a}\hat{e}\hat{f}}
C_{\hat{e}\hat{f}}^{\hat{c}}\right),
\label{Omega1}
\end{equation}
\begin{equation}
\Omega^{\hat{a}}_{(2)}=\frac12\Xi^{\hat{a}}-\frac{\gamma}{\gamma+1}\varepsilon^{\hat{a}\hat{b}\hat{c}}Q_{(\hat{b}\hat{d})}\delta^{\hat{d}\hat{n}}v_{\hat{n}}v_{\hat{c}},
\label{Omega2}
\end{equation}
 where $\gamma=1/\sqrt{1-v^2}$ is a Lorentz factor. 
 
\section{The evolution of a spinning particle in the Kasner universe}

The simplest spatially homogenous and anisotropic universe is that of Bianchi - I type, whose metric is \cite{LL}
\begin{equation}
ds^2=dt^2-a^2(t)(dx^1)^2-b^2(t)(dx^2)^2-c^2(t)(dx^3)^2.
\label{Bianchi-I}
\end{equation}
Comparing this expression with Eq. (\ref{metric}) we have the following expressions  for the nonvanishing elements of the matrix $W_{\hat{a}}^b$:
\begin{equation}
W_{1}^{\hat{1}}=a(t),\ W_{2}^{\hat{2}}=b(t),\ W_{3}^{\hat{3}}=c(t).
\label{Bianchi-I0}
\end{equation}
The elements of the inverse matrix  are 
\begin{equation}
W_{\hat{1}}^1=\frac{1}{a(t)},\ W_{\hat{2}}^{2} = \frac{1}{b(t)},\ W_{\hat{3}}^3=\frac{1}{c(t)}.
\label{Bianchi-I1}
\end{equation} 

As is well known the anholonomity coefficients for the Bianchi-I model are equal to zero.
Hence, $\Upsilon = 0$ too and the ``gravitoelectric'' contribution $\vec{\Omega}_{(1)}$ disappears. 
Then the non-vanishing coefficients of the matrix $Q_{\hat{a}\hat{b}}$ are 
\begin{equation}
Q_{\hat{1}\hat{1}}=-\frac{\dot{a}}{a},\ Q_{\hat{2}\hat{2}}=-\frac{\dot{b}}{b},\ Q_{\hat{3}\hat{3}}=-\frac{\dot{c}}{c}. 
\label{Q1}
\end{equation}
Correspondingly also the vector $\Xi_{\hat{a}}$ disappears.
Finally, the non vanishing components of the ``gravitomagnetic'' contribution to the precession of the Dirac particle in the Bianchi - I universe is, up to cyclic permutations \cite{Kamenshchik:2015iua}
\begin{eqnarray}
&&\Omega_{(2)}^{\hat{1}}=\frac{\gamma}{\gamma+1}v_{\hat{2}}v_{\hat{3}}\left(\frac{\dot{b}}{b}-\frac{\dot{c}}{c}\right).
\label{Bianchi-I2}
\end{eqnarray}

The solution of the Einstein equations for the empty Bianchi-I universe - the Kasner solution\cite{Kasner,Lif-Khal} is 
\begin{equation}
a(t)=a_0t^{p_1},\ b(t)=b_0t^{p_2},\ c(t)=c_0t^{p_3},
\label{Kasner}
\end{equation}
where the Kasner indices $p_1,p_2$ and $p_3$ satisfy the relations
\begin{equation}
p_1+p_2+p_3=1,\ \ p_1^2+p_2^2+p_3^2=1.
\label{Kasner1}
\end{equation}
Correspondingly, Eq. (\ref{Bianchi-I2}) becomes 
\begin{eqnarray}
&&\Omega_{(2)}^{\hat{1}}=\frac{\gamma}{\gamma+1}v_{\hat{2}}v_{\hat{3}}\left(\frac{p_2-p_3}{t}\right)
\label{Bianchi-I4}
\end{eqnarray}
and has some similarity to the Euler equations for rigid body rotation with $p_i$ being corresponding to the moments of inertia.

Obviously, this effect can be essential in the early universe, i.e. at the very small values of the proper cosmic time $t$. 

The Kasner indices $p_1, p_2$ and $p_3$ can be expressed through   
the Lifshitz-Khalatnikov parameter $u$ \cite{Lif-Khal} as 
\begin{equation}
p_1 = -\frac{u}{1+u+u^2},\ p_2=\frac{1+u}{1+u+u^2},\ p_3=\frac{u(1+u)}{1+u+u^2},
\label{Lif-Khal}
\end{equation}
where $u>1$. 

First of all, let us find the velocities of a particle, moving in a Kasner universe, resolving the
the geodesic equations \cite{Harvey,Mashhoon1}.

The geodesic equation for the spatial velocities of a massive particle is 
\begin{equation}
\frac{d^2x^i}{d\tau^2}+2\Gamma^{i}_{j0}\frac{dt}{d\tau}\frac{dx^j}{d\tau}=0.
\label{geod}
\end{equation}
Here $\tau$ is a proper time and the Christoffel symbols are 
\begin{equation}
\Gamma^{1}_{10}=\frac{\dot{a}}{a},\ \Gamma^{2}_{20}=\frac{\dot{b}}{b},\ \Gamma^{3}_{30}=\frac{\dot{c}}{c}. 
\label{geod1}
\end{equation}
The solutions of these equations are 
\begin{equation}
\frac{dx^1}{d\tau}=\frac{C_1}{a^2},\ \frac{dx^2}{d\tau}=\frac{C_2}{b^2},\ \frac{dx^3}{d\tau}=\frac{C_3}{c^2}.
\label{geod2}
\end{equation}
 
 Geodesic equation for the coordinate time parameter $t$ is 
 \begin{equation}
 \frac{d^2 t}{d\tau^2}+\Gamma^0_{ij}\frac{dx^i}{d\tau}\frac{dx^j}{d\tau}=0,
 \label{geod3}
 \end{equation}
 where 
 \begin{equation}
 \Gamma^0_{11}=\dot{a}a,\ \Gamma^0_{22}=\dot{b}b,\ \Gamma^0_{33}=\dot{c}c. 
 \label{geod4}
 \end{equation}
 Substituting expressions (\ref{geod2}) into Eq. (\ref{geod3}) and multiplying it by $\frac{dt}{d\tau}$ we find
 \begin{equation}
 \left(\frac{dt}{d\tau}\right)^2= \frac{C_1^2}{a^2}+\frac{C_2^2}{b^2}+\frac{C_3^2}{c^2}+D,
 \label{geod5}
 \end{equation}
 where $D$ is an integration constant. For the particles in the rest ($C_1=C_2=C_3=0$) the 
 proper time $\tau$ coincides with the coordinate one $t$, hence, $D=1$ and 
 \begin{equation}
 \frac{dt}{d\tau}=\sqrt{1+\frac{C_1^2}{a^2}+\frac{C_2^2}{b^2}+\frac{C_3^2}{c^2}}.
 \label{geod6}
 \end{equation} 
 Let us note, that this is nothing but the factor $\gamma$. 
 Then, the three-velocities are
 \begin{equation}
 v^1=\frac{C_1}{\gamma a^2},\ v^2=\frac{C_2}{\gamma b^2},\ v^3=\frac{C_3}{\gamma c^2}. 
 \label{geod7}
 \end{equation}
 Using the tetrads (\ref{Bianchi-I0}), we can obtain the tetrad velocities
 \begin{equation}
 v^{\hat{1}}=\frac{C_1}{\gamma a},\ v^{\hat{2}}=\frac{C_2}{\gamma b},\ v^{\hat{3}}=\frac{C_3}{\gamma c}. 
 \label{geod8}
 \end{equation}
 
 Substituting the expression (\ref{geod8}) into the expression for the precession velocity 
 (\ref{Bianchi-I2}), we obtain  
 \begin{eqnarray}
 &&\Omega^{\hat{1}}=\frac{1}{\gamma(\gamma+1)}\frac{C_2C_3}{bc}\left(\frac{\dot{b}}{b}-\frac{\dot{c}}{c}\right)\nonumber \\
 &&=\frac{1}{\gamma(\gamma+1)}\frac{C_2C_3(p_2-p_3)t^{p_1-2}}{b_0c_0},
 \label{prec10}
 \end{eqnarray}
 \begin{eqnarray}
 &&\Omega^{\hat{2}}=\frac{1}{\gamma(\gamma+1)}\frac{C_1C_3}{ac}\left(\frac{\dot{c}}{c}-\frac{\dot{a}}{a}\right)\nonumber \\
 &&=\frac{1}{\gamma(\gamma+1)}\frac{C_1C_3(p_3-p_1)t^{p_2-2}}{a_0c_0},
 \label{prec20}
 \end{eqnarray}
 \begin{eqnarray}
 &&\Omega^{\hat{3}}=\frac{1}{\gamma(\gamma+1)}\frac{C_1C_2}{ab}\left(\frac{\dot{a}}{a}-\frac{\dot{b}}{b}\right)\nonumber \\
 &&=\frac{1}{\gamma(\gamma+1)}\frac{C_1C_2(p_1-p_2)t^{p_3-2}}{a_0b_0}.
 \label{prec30}
 \end{eqnarray}
 Let us suppose that $p_1 \leq p_2 \leq p_3$ and can be parametrised as in Eq. 
 (\ref{Lif-Khal}). 
 The dependence of the factor $\gamma$ on time (see Eq. (\ref{geod6}) is rather involved, which makes the precession equations unsolvable in simple terms. Thus, let us consider the 
 Kasner universe close to the singularity when $t \rightarrow 0$. In this case 
 \begin{equation}
 \gamma = \frac{C^3}{t^{p_3}}.
 \label{gamma1}
 \end{equation}
 Substituting this expression into Eqs. (\ref{prec10})--(\ref{prec30}), we obtain
 \begin{equation}
 \Omega^{\hat{1}}=\frac{C_2C_3(p_2-p_3)t^{2p_3+p_1-2}}{C_3^2b_0c_0},
 \label{prec40}
 \end{equation}
 \begin{equation}
 \Omega^{\hat{2}}=\frac{C_1C_3(p_3-p_1)t^{2p_3+p_2-2}}{C_3^2a_0c_0},
 \label{prec50}
 \end{equation}
 \begin{equation}
 \Omega^{\hat{3}}=\frac{C_1C_2(p_1-p_2)t^{3p_3-2}}{C_3^2a_0b_0}.
 \label{prec60}
 \end{equation}
 Writing down explicitly the exponents 
 \begin{eqnarray}
 &&2p_3+p_1-2=\frac{-u-2}{1+u+u^2},\nonumber \\
 &&2p_3+p_2-2=\frac{u-1}{1+u+u^2},\nonumber \\
 &&3p_3-2=\frac{u^2+u-2}{1+u+u^2},
 \label{degrees}
 \end{eqnarray}
  we see that only $\Omega^{\hat{1}}$ survives in the vicinity of singularity.
  Thus, the equations for the spin precession (\ref{prec}) take the form
  \begin{eqnarray}
  &&\dot{s}^{\hat{2}}=Et^{q}s^{\hat{3}},\nonumber \\
  &&\dot{s}^{\hat{3}}=-Et^{q}s^{\hat{2}},
  \label{degrees1}
  \end{eqnarray}
  where 
  \begin{equation}
  E = \frac{C_2(u^2-1)}{C_3b_0c_0(1+u+u^2)} > 0,
  \label{E}
 \end{equation}
 \begin{equation}
 q = \frac{-u-1}{1+u+u^2}.
 \label{q}
 \end{equation}
 It is easy to integrate this couple of equations and to find that 
 \begin{eqnarray}
 &&s^{\hat{2}} = s_0\sin\left(\frac{Et^{q+1}}{q+1}+\varphi\right),\nonumber \\
 &&s^{\hat{3}} = s_0\sin\left(\frac{Et^{q+1}}{q+1}+\varphi\right),
 \label{q1}
 \end{eqnarray}
 where the amplitude $s_0$ and the phase $\varphi$ are the integration constants.
 It is important that
 $$
 q+1 = \frac{u^2}{1+u+u^2} > 0.
 $$
 That means that in spite of the fact that the precession velocity tends to infinity,
 this singularity is an integrable one and the solution of the equations of the precession 
 shows that the average spin vector $\vec{s}$ tends to some fixed direction (determined by the phase $\varphi$) in quite a regular way. 
 
Let us consider an opposite limiting case of the infinite expansion of the universe $t \rightarrow \infty$. This limit was considered in paper \cite{Mashhoon1}, where it was related to the possible production of cosmic jets. In this context, the component velocity of a particle    
oriented along the axis of contracting dimension tends to the velocity of light. 
What is the behavior of the jet angular momentum? It is easy to see that in this case the factor $\gamma$ behaves as 
\begin{equation}  
\gamma = \frac{C^1}{t^{p_1}}.
 \label{gamma2}
 \end{equation} 
 Then, the components of the precession velocity are 
 \begin{equation}
 \Omega^{\hat{1}}=\frac{C_2C_3(p_2-p_3)t^{3p_1-2}}{C_1^2b_0c_0},
 \label{prec400}
 \end{equation}
 \begin{equation}
 \Omega^{\hat{2}}=\frac{C_1C_3(p_3-p_1)t^{2p_1+p_2-2}}{C_3^2a_0c_0},
 \label{prec500}
 \end{equation}
 \begin{equation}
 \Omega^{\hat{3}}=\frac{C_1C_2(p_1-p_2)t^{2p_1+p_3-2}}{C_3^2a_0b_0}.
 \label{prec600}
 \end{equation}
 The largest power of $t$ is
 $$
 2p_1+p_3-2=-\frac{u^2+3u+2}{1+u+u^2} < 0.
 $$
 Thus, all the components of the precession velocity tend rapidly to zero when $t \rightarrow 
 \infty$. 
 
 Let us recall that due to the equivalence principle, the macroscopic angular momentum is evolving like spin. So, the angular momenta of jets are changing very slowly. 
 
 It is curious, that even in the vicinity of the cosmological singularity of the Kasner universe, the angular momentum remains quite stable and does not exhibit any singular behavior. 
 Perhaps, one can say that, in a way, the rotation possesses some smoothing effect. 
 
\section{Geodesics and jets in the Heckmann-Schucking universe} 

In paper \cite{Mashhoon1} an interesting possibility of production of jets in the Kasner spacetime was considered. Such a possibility is connected with the fact that 
at the expansion the velocity of test particles in the contracting direction is growing tending to that of light (see Eq. (\ref{geod7}). It is particularly interesting to study such e phenomenon in a more realistic Heckmann--Schucking model, which represents the Bianchi-I universe filled with dust \cite{Heck-Schuck}. Note that this solution can be easily generalised for the case when the stiff matter and the cosmological constant are also present \cite{Khal-Kam,Khal-Kam1}.  

Now, let us give some details of the Heckmann-Schucking solution following the approach developed in \cite{Khal-Kam}.
It is convenient to represent the scale factors $a,b$ and $c$ for a Bianchi-I universe
(\ref{Bianchi-I}) as 
\begin{eqnarray}
&&a(t)=R(t)\exp(\alpha(t)+\beta(t)),\nonumber \\
&&b(t)=R(t)\exp(\alpha(t)-\beta(t)),\nonumber \\
&&c(t)=R(t)\exp(-2\alpha(t)).
\label{HS}
\end{eqnarray}
The Friedmann-type equation for the function $R(t)$ can be written in the form:
\begin{equation}
\frac{\dot{R}^2}{R^2}=\dot{\alpha}^2+\frac{\dot{\beta}^2}{3}+\frac{M}{R^3},
\label{HS1}
\end{equation}
where the term $\frac{M}{R^3}$ is related to the presence of dust in the universe, while 
the squared derivatives of anisotropy factors $\alpha(t)$ and $\beta(t)$ represent the influence of the anisotropy on the dynamics of the scale factor $R(t)$.
The former, in turn satisfy the equations 
\begin{eqnarray}
&&\dot{\alpha}=\frac{\alpha_0}{R^3},\nonumber \\
&&\dot{\beta}=\frac{\beta_0}{R^3},
\label{HS2}
\end{eqnarray}
where $\alpha_0$ and $\beta_0$ are the integrations constants. 
Then
\begin{equation}
\frac{\dot{R}^2}{R^2}=\frac{M}{R^3} +\frac{S}{R^6},
\label{HS3}
\end{equation}
where 
\begin{equation}
S=\alpha_0^2+\frac{\beta_0^2}{3}.
\label{HS4}
\end{equation}
 Eq. (\ref{HS3})  can be easily integrated, giving
\begin{equation}
 R(t) = \left(\frac94Mt^2+3\sqrt{S}t\right)^{1/3},
 \label{HS5}
 \end{equation}
 where we have chosen initial conditions in such a way that  
 at the moment $t=0$ the universe has a cosmological singularity.
 Substituting this expression into Eqs. (\ref{HS2}) we have 
 \begin{equation}
 \alpha(t) = \alpha_0\int \frac{dt}{\frac94Mt^2+3\sqrt{S}t} = \frac{\alpha_0}{3\sqrt{S}}
 \ln\left(\frac{t}{t+\frac{4\sqrt{S}}{3M}}\right),
 \label{HS6}
 \end{equation}
 where the integration constant is chosen in such a way that  $\alpha(t) \rightarrow 0$   
when $t \to \infty$. The expression for $\beta(t)$ is quite analogous. Now, we can write down 
the expressions for the scale factors of the Bianchi-I universe:
\begin{eqnarray}
&&a(t)=\left(\frac94M\right)^{\frac13}t^{p_1}(t+t_0)^{\frac23-p_1},\nonumber \\
&&b(t)=\left(\frac94M\right)^{\frac13}t^{p_2}(t+t_0)^{\frac23-p_2},\nonumber \\ 
&&c(t)= \left(\frac94M\right)^{\frac13}t^{p_3}(t+t_0)^{\frac23-p_3},
\label{HS7}
\end{eqnarray}
where 
\begin{equation}
t_0 = \frac{4\sqrt{S}}{3M},
\label{HS8}
\end{equation}
and the exponents 
\begin{eqnarray}
&&p_1=\frac13+\frac{\alpha_0+\beta_0}{3\sqrt{S}},\nonumber \\
&&p_2=\frac13+\frac{\alpha_0-\beta_0}{3\sqrt{S}},\nonumber\\
&&p_3=\frac13+\frac{-2\alpha_0}{3\sqrt{S}}
\label{HS9}
\end{eqnarray}
satisfy the Kasner relations (\ref{Kasner1}). It is easy to see that in the vicinity of the singularity when $t \ll t_0$ the solution (\ref{HS9}) behaves as the Kasner one  (\ref{Kasner}), while at $t \gg t_0$ the scale factors behave as 
\begin{equation}
a(t)\sim t^{2/3},\ b(t) \sim t^{2/3},\ c(t) \sim t^{2/3},
\label{HS10}
\end{equation}
i.e. the universe behaves as a Friedmann flat universe filled with dust. The magnitude $t_0$ characterises the time scale where the regime is changed.

Let us note that we cannot make a naive transition from the Heckmann-Schucking solution  
(\ref{HS7}) to the Kasner solution (\ref{Kasner}) by requiring that $M \to 0$. It is connected with the fact that the integrals for the anistropy functions (\ref{HS6}) logarithmically diverge at $t \to \infty$  if $M=0$. Thus, the scale $t_0=4\sqrt{S}/3M$ looses the sense and one should introduce another time scale to make it convergent. This phenomenon can be considered as a particular example of the automodelity of the second order \cite{Baren}.
Under the automodelity of the second order one means the nonexistence of a finite limit of some observable when a particular parameter is tending to zero or to infinity. Instead, one has a power dependence on this parameter.  We would like to stress that this power is similar to anomalous dimension in the renormalization group approach to quantum field theory \cite{BS}.

An interesting feature of the Heckmann-Schucking solution (\ref{HS7}) is that the scale factor 
$a(t)$ has a non-monotonic behaviour if, as usual, we choose $p_1 < 0$. Indeed, this factor is infinitely large at the singularity at $t = 0$, then it begin decreasing arriving the point of minimal contraction at the moment 
\begin{equation}
t_{\rm min} = -\frac{3p_1}{2}t_0=\frac{3u}{2(1+u+u^2)}t_0.
\label{mini}
\end{equation}
The minimal value of the scale factor $a$ is 
\begin{eqnarray}
 &&a_{\rm min}=\left(\frac{4S}{M}\right)^{\frac13}\times f(u),\nonumber \\
 &&f(u)=\frac{
 (3u)^{-\frac{u}{1+u+u^2}}
 (2u^2+5u+2)^{\frac{2u^2+5u+2}{3(1+u+u^2)}}}
 {[2(1+u+u^2)]^{\frac23}}.
 \label{mini1}
 \end{eqnarray}
  \begin{figure}[h]
\centerline{\epsfxsize 8cm \epsfbox{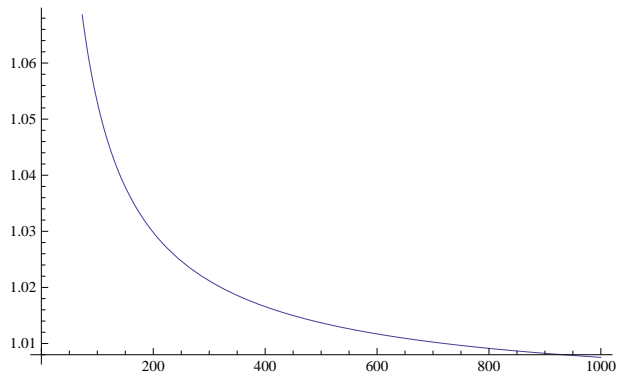}}
\caption{\small The dependence of the function $f(u)$ on the Lifshitz - Khalatnikov parameter 
$u$.
 \label{Fig.1}}
\end{figure}
The plot of the function $f(u)$ is presented in Fig. 1. It decreases monotonically from $f(1)\approx 1.9$ until $f(\infty)=1$. 

Now, it is convenient to  write down  the explicit expression for the component of the tetrad velocity oriented along the contracting-expanding axis 1:
\begin{equation}
v^{\hat{1}}=\frac{C_1}{a\sqrt{1+\frac{C_1^2}{a^2}+\frac{C_2^2}{b^2}+\frac{C_3^2}{c^2}}}.
\label{rapid}
\end{equation}
This component is maximal for the particles with vanishing transverse velocities, i.e. with 
$C_2 =C_3=0$. Then, for a fixed value of $C_1$, the maximal velocity is 
\begin{equation}
v^{\hat{1}}_{\rm max}=\frac{C_1}{\sqrt{a^{2}_{\rm min}+C_1^2}}.
\label{rapid1}
\end{equation}
It is approaching velocity of light when $C_1$ is growing.

\section{Precession in a Bianchi-IX universe}  

The matrix $W^{\hat{b}}_{\ a}$ for the Bianchi - IX metric (see e.g. \cite{Ryan}) can be written as 
\begin{equation}
W^{\hat{b}}_{\ a}=\left(\begin{array}{ccc}
-a\sin x^3&a\sin x^1\cos x^3&0\\
b\cos x^3&b\sin x^1\sin x^3&0\\
0&c\cos x^1&c
\end{array}
\right),
\label{9}
\end{equation}
where $a, b$ and $c$ are some functions of time  as usual. 
Its inverse matrix $W^c_{\ \hat{b}}$ is 
\begin{equation}
W^c_{\ \hat{b}}=\left(
\begin{array}{ccc}
-\frac{1}{a}\sin x^3&\frac{1}{b}\cos x^3&0\\
\frac{1}{a}\frac{\cos x^3}{\sin x^1}&\frac{1}{b}\frac{\sin x^3}{\sin x^1}&0\\
-\frac{1}{a}\frac{\cos x^1\cos x^3}{\sin x^1}&-\frac{1}{b}\frac{\sin x^3\cos x^1}{\sin x^1}&\frac{1}{c}
\end{array}
\right).
\label{91}
\end{equation}
The nonvanishing anholonomity coefficients are 
\begin{equation}
C_{\hat{1}\hat{2}}^{\hat{3}}=\frac{c}{ab},\ {\rm and\ cyclic\ permutations}.
\label{92}
\end{equation}
 Then
 \begin{equation}
 \Upsilon = 2\left(\frac{c}{ab}+\frac{b}{ac}+\frac{a}{bc}\right).
 \label{93}
 \end{equation}
The nonvanishing coefficients of the matrix $Q_{\hat{a}\hat{b}}$ are the same as in Eq. (\ref{Q1}). Hence, $\Xi_{\hat{a}}$ is again equal to zero. 
The ``gravitoelectric'' precession velocity, up to cyclic permutations,  is
\begin{eqnarray}  
&&\Omega^{\hat{1}}_{(1)}=v^{\hat{1}}\left(\frac{c}{ab}+\frac{b}{ac}-\frac{a}{bc}\right),
\label{94}
\end{eqnarray}

The expressions for the components of the ``gravitomagnetic'' velocity  
$\vec{\Omega}_{(2)}$ 
remain the same as in Bianchi-I universe (see Eq. (\ref{Bianchi-I2})).

Thus, we have seen that the ``gravitomagnetic'' velocity in the Bianchi-IX universe is the same as in Bianchi-I universe, however, in the Bianchi-IX universe there is also the ``gravitoelectric'' precession. The presence of this term (\ref{94}) is connected with the presence of a spatial curvature in the Bianchi-IX universe, in contrast to the Bianchi-I universe. It is connected with the fact that the anholonomity coefficients are nonvanishing in the Bianchi-IX universe.    

Now, let us discuss what happens with the precession of the Dirac particles in the Bianchi-IX universe, evolving towards the cosmological  singularity. As was discovered at the end of sixties the Bianchi-IX universe approaches the singularity in an oscillating way \cite{BKL,Misner} and these oscillations have chaotic character \cite{chaos}.  The evolution towards the  singularity can be described by the subsequence of the periods when the universe behaves like a Kasner universe (\ref{Kasner}), (\ref{Kasner1}), separated by time intervals when one Kasner regime is substituted by another one.  Let us remind how these changes occur. The Kasner indices $p_1, p_2$ and $p_3$ can be expressed through   
the Lifshitz-Khalatnikov parameter $u$ \cite{Lif-Khal} as 
\begin{equation}
p_1 = -\frac{u}{1+u+u^2},\ p_2=\frac{1+u}{1+u+u^2},\ p_3=\frac{u(1+u)}{1+u+u^2},
\label{Lif-Khala}
\end{equation}
where $u>1$. 
The perturbative terms in the Einstein equations, connected with the spatial curvature, induce the transition to another Kasner regime (which is called ``epoch'' \cite{LL,BKL}). Such that 
\begin{equation}
p'_1=p_2(u-1),\ p'_2 = p_1(u-1),\ p'_3 = p_3(u-1).
\label{Lif-Khal1}
\end{equation}
That means that  if during  the preceding epoch, the universe is expanding along the first axis and contracting along the second and third axes, in the successive epoch it begins expanding along the second axis, i.e. the first and second axis change their roles. There is another type of transition when the parameter $u$ becomes less than 1. In this case the change of the ``Kasner era'' occurs \cite{LL,BKL}. This change is described by the following formula:
\begin{equation}
p'_1=p_1\left(\frac{1}{u}\right),\ p'_2=p_3\left(\frac{1}{u}\right),\ p'_3=p_2\left(\frac{1}{u}\right).
\label{Lif-Khal2}
\end{equation}
The transition from one Kasner era to another  can be described by the mapping transformation of the interval $[0,1]$ into itself by the formula 
\begin{equation}
 Tx = \left\{\frac{1}{x}\right\},\ \ x_{s+1} = \left\{\frac{1}{x_{s}}\right\},
 \label{Gauss0}
 \end{equation}
where curly brackets stay for the fractional part of a number. 
This transformation belongs to the so-called expanding transformations
of the interval $[0, 1]$, i.e., transformations $x\sim f(x)$ with $|f'(x)| > 1$. Such
transformations possess the property of exponential instability: if we take
initially two close points, their mutual distance increases exponentially under
the iterations of the transformations. It is well known that the exponential
instability leads to the appearance of strong stochastic properties \cite{chaos}.

Now, let us describe what happens with our angolar velocities $\vec{\Omega}_{(1)}$
and $\vec{\Omega}_{(2)}$ when the universe oscillating approach to the singularity.
The expression for $\vec{\Omega}_{(2)}$ can be written as 
\begin{eqnarray}
&&\Omega^{\hat{1}}_{(2)}=\frac{\gamma}{(\gamma+1)t}v_{\hat{2}}v_{\hat{3}}\cdot\frac{1-u^2}{1+u+u^2},\nonumber \\
 &&\Omega^{\hat{2}}_{(2)}=\frac{\gamma}{(\gamma+1)t}v_{\hat{1}}v_{\hat{3}}\cdot\frac{2u+u^2}{1+u+u^2},\nonumber \\
 &&\Omega^{\hat{3}}_{(2)}=-\frac{\gamma}{(\gamma+1)t}v_{\hat{1}}v_{\hat{2}}\cdot\frac{1+2u}{1+u+u^2}.
 \label{flip}
 \end{eqnarray}
 It is easy to show that after the change of the Kasner epoch the new expressions for the components of the velocity can be obtained by substitution  $u \rightarrow -u$ in this equation. It means that the component the first component does not change the sign, the third component changes the sign while the second component changes the sign if $u > 2$. 
 
 After the change of the Kasner era all the components of the velocity $\vec{\Omega}_{(2)}$
 just change the sign, preserving the absolute values,  as it follows immediately from (\ref{Lif-Khal2}).   
 
 The leading terms for the components of the velocity  $\vec{\Omega}_{(1)}$ 
 are
 \begin{eqnarray}
 &&\Omega^{\hat{1}}_{(1)} \sim -v^{\hat{1}}(t)^{\left(-1-\frac{2u}{1+u+u^2}\right)},\nonumber \\ 
&&\Omega^{\hat{b}}_{(1)} \sim v^{\hat{b}}(t)^{\left(-1-\frac{2u}{1+u+u^2}\right)} ,\   b=2, 3.\label{flip2}
  \end{eqnarray}
The change of epochs boils down to
 \begin{eqnarray}
 &&\Omega^{\hat{2}}_{(1)} \sim -v^{\hat{2}}(t)^{\left(-1-\frac{2u-2}{1-u+u^2}\right)},\nonumber \\ 
&&\Omega^{\hat{a}}_{(1)} \sim v^{\hat{a}}(t)^{\left(-1-\frac{2u-2}{1-u+u^2}\right)} ,\   a=1, 3.
\label{flip2a}
  \end{eqnarray}
  Curiously, the change of eras leave leading terms under consideration
   intact.
  Thus, we have seen, that the precession of the Dirac particle in the Bianchi-IX universe 
  evolving towards the singularity also follows a chaotic pattern.
  
  \section{Discussion and Outlook}  

We have seen that the precession of a Dirac particle spin exists already  in the Bianchi-I universe. Interestingly, the  Kasner indices play the role similar to the moments of inertia in the Euler equation for the rigid body precession.  
In the Bianchi-IX universe the precession  acquires the chaotic character due to the stochasticity of the oscillatory approach to the cosmological singularity \cite{BKL,Misner}. Remarkably, the formulae for the changes of the precession direction are nicely expressible in terms of the     
Lifshitz-Khalatnikov parameter $u$. 

What physical consequences could it have for the very early universe?

Let us note first that precession due to anisotropy of the Universe may be considered as generated by some effective magnetic field. The latter may be easily obtained by equating the angular velocity to that of Larmor precession.
For the definiteness, in Bianchi-I universe it reads, up to cyclic permutations
\begin{eqnarray}
&&H^{\hat{1}}=\frac{m \gamma}{2eg(\gamma+1)}v_{\hat{2}}v_{\hat{3}}\left(\frac{p_2-p_3}{t}\right)
\label{H}
\end{eqnarray}
As a result, the anisotropy of the universe provides all the Dirac particles with effective anomalous magnetic moments. In particular, the transitions between Dirac neutrinos and their sterile partners may be induced in such a way.
Moreover, due to equivalence principle, these conclusions may be extended to particles of any spin \cite{Teryaev:1999su} and also to classical rotators  \cite{ter-ob-sil1}. The latter fact opens the possibility to study the role of the discussed precession effects for the formation of structures in the very early universes and angular momentum of cosmic strings \cite{Mashhoon}. 

Also, equivalence principle leads  to the precession frequencies of spin and velocity differing by factor 2, so that the helicity is conserved in the non-inertial frame rotating with the same frequency but it is flipped \cite{Teryaev:1999su} in the inertial frame. This effect is especially interesting for massive Dirac neutrinos.
If they are produced in the very early Universe as active ones, the gravity-induced helicity flip may turn them 
to sterile neutrinos which remain in this state after Universe becomes isotropic and contribute to fermionic dark matter. As soon as the rotation period is defined by the age of the Universe in the anisotropic phase, the 
amounts of sterile and active neutrinos at the end of this phase are of the same order:
\begin{equation}    
\frac{N_{\it sterile}}{N_{\it active}} \sim 1.
\end{equation}
If the spin happens to perform the rotation for an angle close to $\pi$, the velocity will rotate for the angle close to $2\pi$ 
and the most of the fermions will become sterile:
\begin{equation}    
\frac{N_{\it sterile}}{N_{\it active}} \gg 1.
\label{tune}
\end{equation}
This opens \cite{Kamenshchik:2015iua}, in principle, the possibility to attribute the dark matter to the contribution of light sterile neutrinos whose abundance would be much larger than that of thermal ones. The validity of (\ref{tune}) would require a sort of fine tuning, but not too strong one, as the required 
excess of sterile neutrinos is about two orders of magnitude, and the closeness of the rotation angle to $\pi$ should be also at the percent level.   

The anisotropic metrics, in the case of some scale parameters being much smaller than others may provide the model of transitions between spaces of different (effective) dimension \cite{Fiziev:2011qm,Stojkovic:2014lha,Afshordi:2014cia}. The spin dynamics in that case is manifesting the interesting effects \cite{Silenko:2013kva}. 
Other interesting directions of investigation could be connected with the study of the spin precession in Bianchi II universes, in the generalized Melvin cosmologies in the presence of electromagnetic fields \cite{Jennie} and in the double Kasner universes \cite{Mashhoon}. 
We hope to study this topics in detail in future publications.

\acknowledgments
The authors are grateful to  I.M. Khalatnikov, Yu.N. Obukhov, V.A. Rubakov  and A.J. Silenko for useful discussions. The work of A.K. was partially supported by the RFBR through the grant No 14-02-00894 while the work of O.T. was partially supported through the grant No 14-01-00647.  O.T. is grateful to the Sezione di INFN in Bologna and to the Dipartimento di Fisica e Astronomia dell'Universit\`a di Bologna for kind hospitality during his visits to Bologna in the Falls of 2014 and 2015.

\end{document}